\documentstyle[aps,eqsecnum]{revtex}

\begin{document}
\title{The imposition of Cauchy data to the Teukolsky equation III:
 The rotating case}
\author{Manuela Campanelli$^1$ , Carlos O. Lousto$^1$, John Baker$^2$,
Gaurav Khanna$^2$, and Jorge Pullin$^2$}
\address{1. Instituto de Astronom\'\i a y F\'\i sica del Espacio,\\
Casilla de Correo 67, Sucursal 28\\
(1428) Buenos Aires, Argentina}
\address{2. Center for Gravitational Physics and Geometry,
Department of Physics, The Pennsylvania State University,\\
104 Davey Lab, University Park, PA 16802}
\date{\today }
\maketitle

\begin{abstract}
We solve the problem of expressing the Weyl scalars $\psi $ that describe
gravitational perturbations of a Kerr black hole in terms of Cauchy data. To
do so we use geometrical identities (like the Gauss-Codazzi relations) as
well as Einstein equations. We are able to explicitly express $\psi $ and $%
\partial _t\psi $ as functions only of the extrinsic curvature and the
three-metric (and geometrical objects built out of it) of a generic
spacelike slice of the spacetime. These results provide the link between
initial data and $\psi $ to be evolved by the Teukolsky equation, and can be
used to compute the gravitational radiation generated by two {\it orbiting }
black holes in the close limit approximation. They can also be used to
extract waveforms from spacetimes completely generated by numerical methods.
\end{abstract}

\section{Introduction}

In Ref. \cite{CL97} the question was raised of how to impose initial data to
the Teukolsky equation (that describe perturbations around a rotating black
hole). We noted that the expressions of Chrzanowsky\cite{C75} for the Weyl
scalars $\psi _4$ and $\psi _0$ in terms of metric perturbations were
written as second order operators on the four-metric and appeared
inconvenient at the moment to use them for building up the initial values
needed to start the integration of the Teukolsky equation.

The work of reference \cite{CL97} showed how to solve the problem for a
nonrotating background, i.e. perturbations around a Schwarzschild hole by
relating Weyl scalars $\psi ,$ to the Moncrief waveforms $\phi _M$, an
alternative description of metric perturbations explicitly built up out of
the three-metric $\stackrel{\_}{g}_{ij}$ and the extrinsic curvature $K_{ij}$
of the hypersurface $t=$ constant. In Ref. \cite{CKL98} the $\psi $ -- $\phi
_M$ relations were successfully tested with a program for integration of the
Teukolsky equation.

It is not obvious how to extend the above techniques to the rotating case.
Thus, in the present paper we turned to a more geometrical approach that
lead us to the desired relations for {\it rotating }holes. In Sec. II we
collect the results of the 3+1 decomposition reviewed in Ref. \cite{AABY96}
relevant for our derivation. This has the advantage that makes $\psi $ to be
automatically independent of the shift, so our task is reduced to prove that
terms depending on the first perturbative order lapse vanish. This is made
in Sec. III, where we also build up $\partial _t\psi $ in terms only of $%
\stackrel{\_}{g}_{ij}$and $K_{ij}.$ This results allow to compare, given the
initial data, evolution through integration of the full Einstein equations
and Teukolsky equation (linearization around a Kerr hole), and test, for
instance, the close limit approximation for orbiting holes.

Notation: We use Ref. \cite{MTW73} conventions. An overbar on geometric
quantities means that they are three-dimensional quantities, i.e. defined on
the $t=$ constant hypersurfaces $\Sigma _t$ (an exception to this rule is
the complex conjugation of the vector $m^\alpha ,$ i.e. $\stackrel{\_\_}{m}%
^\alpha $). $(\alpha ,\beta )$ and $[\alpha ,\beta ]$ on indices $\alpha
,\beta $ represent the usual symmetric and antisymmetric parts respectively.
Greek letters indices run from 0 to 3 while latin letters indices run from 1
to 3. Subindexes (0) and (1) mean pieces of exclusively zeroth and first
order respectively.

\section{Geometric structure and gravitation}

Following Ref. \cite{AABY96} we write the metric as 
\begin{equation}
ds^2=-N^2(\theta ^0)^2+g_{ij}\theta ^i\theta ^j,
\end{equation}
with $\theta ^0=dt$ and $\theta ^i=dx^i+N^idt$, where $N^i$ is the shift
vector and $N$ the lapse.

The cobasis $\theta ^\alpha $ satisfies 
\begin{equation}
d\theta ^\alpha =-{\frac 12}C_{\beta \gamma }^\alpha \theta ^\beta \wedge
\theta ^\gamma
\end{equation}
with $C_{0j}^i=-C_{j0}^i=\partial _jN^i$ and all other structure
coefficients zero. Note that $\bar g_{ij}=g_{ij}$ and $\bar g^{ij}=g^{ij}$.

The spacetime connection one-forms are defined by 
\begin{equation}
\omega _{\beta \gamma }^\alpha =\Gamma _{\beta \gamma }^\alpha +g^{\alpha
\delta }C_{\delta (\beta }^\epsilon g_{\gamma )\epsilon }-{\frac 12}C_{\beta
\gamma }^\alpha =\omega _{(\beta \gamma )}^\alpha +\omega _{[\beta \gamma
]}^\alpha ,
\end{equation}
where $\Gamma _{\beta \gamma }^\alpha $ denotes the Christoffel symbol.
These connection forms are written out explicitly in \cite{ABY97}. In
particular, ${\omega }^i{}_{jk}={\Gamma }^i{}_{jk}=\bar {{\Gamma }}^i{}_{jk}$
, and the extrinsic curvature is given by 
\begin{equation}
K_{ij}=-N\omega _{ij}^0\equiv -{\frac 12}N^{-1}\widehat{\partial }_0g_{ij},
\label{curvextr}
\end{equation}
where we define the operator 
\begin{equation}
\widehat{\partial }_0={\frac \partial {\partial t}}-{\cal L}_{{\bf N}},
\end{equation}
with ${\cal L}_{{\bf N}}$ the Lie derivative on the hypersurface $\Sigma _t$
with respect to the vector $N^i$. Note that $\widehat{\partial }_0$ and $%
\partial _i$ commute.

The Riemann curvature tensor is given by\cite{ABY97} 
\begin{equation}
R^\alpha {}_{{\beta }\rho {\sigma }}={\partial }_\rho {\omega }_{{\beta }{%
\sigma }}^\alpha -{\partial }_\sigma {\omega }^{{\alpha }}{}_{{\beta }\rho }+%
{\omega }^\alpha {}_{{\lambda }\rho }{\omega }^\lambda {}_{{\beta }{\sigma }%
}-{\omega }^\alpha {}_{{\lambda }{\sigma }}{\omega }^\lambda {}_{{\beta }%
\rho }-{\omega }^\alpha {}_{{\beta }{\lambda }}C^{{\lambda }}{}_{\rho {%
\sigma }}  \label{Riemann}
\end{equation}

For rewriting in the next section the Weyl scalars in terms of hypersurface
quantities only, we relate the spacetime Riemann tensor components to the
3-dimensional Riemann and the extrinsic curvature tensors 
\begin{eqnarray}
R_{ijkl} &=&\bar R_{ijkl}+2K_{i[k}K_{l]j},  \label{rijkl} \\
R_{0ijk} &=&2N\bar \nabla _{[j}K_{k]i},  \label{r0jkl} \\
R_{0i0j} &=&N(\widehat{\partial }_0K_{ij}+NK_{ip}K^p{}_j+\bar \nabla _i\bar 
\nabla _jN).  \label{r0j0l}
\end{eqnarray}
Another important relation in three dimensions is 
\begin{equation}
\bar R_{ijkl}=2g_{i[k}\bar R_{l]j}+2g_{j[l}\bar R_{k]i}+\bar Rg_{i[l}g_{k]j}.
\label{r3ijkl}
\end{equation}
The Ricci tensor $R_{\alpha \beta }=R^\sigma {}_{\alpha \sigma \beta }$ is
given by 
\begin{eqnarray}
R_{ij} &=&\bar R_{ij}-N^{-1}\widehat{\partial }%
_0K_{ij}+KK_{ij}-2K_{ip}K^p{}_j-N^{-1}\bar \nabla _i\bar \nabla _jN,
\label{rij} \\
R_{0i} &=&N\bar \nabla ^j(Kg_{ij}-K_{ij}),  \label{r0i} \\
R_{00} &=&N\stackrel{\_}{\nabla }^2N-N^2K_{pq}K^{pq}+N\widehat{\partial }_0K.
\label{r00}
\end{eqnarray}
In order to incorporate the source terms we consider the Einstein equations
as $R_{\alpha \beta }=T_{\alpha \beta }-{\frac 12}g_{\alpha \beta }T$. For
instance, the ``Energy constraint'' is defined by 
\begin{equation}
G^0{}_0={\frac 12}(K_{mk}K^{mk}-K^2-\bar R)=T^0{}_0.  \label{G00}
\end{equation}
Finally, from its definitions 
\begin{equation}
\widehat{\partial }_0\bar R_{ij}=\bar \nabla _k(\widehat{\partial }_0\bar 
\Gamma _{ij}^k)-\bar \nabla _j(\widehat{\partial }_0\bar \Gamma _{ik}^k),
\label{Rijpunto}
\end{equation}
where 
\begin{equation}
\widehat{\partial }_0\bar \Gamma _{ij}^k=-2\bar \nabla _{(i}(NK_{j)}{}^k)+%
\bar \nabla ^k(NK_{ij}).  \label{gamapunto}
\end{equation}

Note that writing equations in terms of $\widehat{\partial }_0$ instead of $%
\partial _t$ allowed us to get rid of the shift dependence. This is because $%
\widehat{\partial }_0$ is orthogonal to the spacelike hypersurface $\Sigma
_t.$

\section{Weyl scalars for Kerr perturbations}

For the computation of gravitation radiation from astrophysical sources it
is convenient to work with the Weyl scalar

\[
\psi _4=-C_{\alpha \beta \gamma \delta }n^\alpha \overline{m}^\beta n^\gamma 
\overline{m}^\delta , 
\]

since it is directly related to the outgoing gravitational waves. For
perturbations around a Kerr hole we have 
\[
-\psi _4=R_{ijkl}n^i\overline{m}^jn^k\overline{m}^l+4R_{0jkl}n^{[0}\overline{%
m}^{j]}n^k\overline{m}^l+4R_{0j0l}n^{[0}\overline{m}^{j]}n^{[0}\overline{m}%
^{l]}. 
\]

Eqs. (\ref{rijkl}) and (\ref{r0jkl}) directly give us the two first terms in
the above sum in terms of hypersurface geometrical objects $(g_{ij},$ $%
K_{ij}).$ In the last term we have to make use of Einstein equation (\ref
{rij}) to eliminate $\widehat{\partial }_0K_{ij}.$ If one now considers
first order perturbations around a Kerr hole, one would have to consider in $%
\psi _4$ two types of terms: terms that involve first order perturbative
Riemann tensors contracted with the background tetrads and terms that
involve the Riemann tensor of the background contracted with three
background and one perturbative tetrads. It is not difficult to see that the
latter terms vanish for the Kerr background. For the Kerr geometry the only
non-vanishing Weyl scalar is $\psi _2=R_{\alpha \beta \gamma \delta
}l^\alpha m^\beta n^\gamma \overline{m}^\delta $ and one can quickly see
that the above contributions, even with one of the tetrads being a
perturbative one, still vanish. For instance, consider the term $%
R_{ijkl}{}_0n_{(1)}^i\overline{m}^jn^k\overline{m}^l$. This term vanishes
because it is contracted with two $\overline{m}$ vectors, and any
contraction with a repeated tetrad vector of the Riemann tensor vanishes for
the Kerr spacetime. Similar arguments apply to the other terms.

Let us turn our attention to the terms that involve the first order Riemann
tensors contracted with the background tetrads. Taking a look at equations (%
\ref{rijkl})-(\ref{r0j0l}) we see that if one considers first order
perturbations, we will have expressions involving the first order extrinsic
curvature, metric, and lapse. We do not want our final expression to depend
on the perturbative lapse. It is easy to see that it actually does not
depend on it. For $R_{0ijk}$ we see that the lapse appears as an overall
factor. So the expression evaluated for the perturbative lapse is
proportional to the expression evaluated in the background, which vanishes.
For $R_{0i0j}$ if we rewrite it using the Einstein equation (\ref{rij})
again the lapse appears as an overall factor and the same argument as for $%
R_{0ijk}$ applies. As a separate check, we have verified the independence on
the perturbative lapse and shift using computer algebra.

The final result for the first order expansion of the Weyl scalar $\psi _4$
therefore is, 
\begin{eqnarray}
-\psi _4 &=&\left[ \stackrel{\_}{R}_{ijkl}+2K_{i[k}K_{l]j}\right] _{(1)}n^i%
\overline{m}^jn^k\overline{m}^l-4N_{(0)}\left[ K_{j[k,l]}+\stackrel{\_}{%
\Gamma }_{j[k}^pK_{l]p}\right] _{(1)}n^{[0}\overline{m}^{j]}n^k\overline{m}^l
\label{psi} \\
&&\ +4N_{(0)}^2\left[ \stackrel{\_}{R}_{jl}-K_{jp}K_l^p+KK_{jl}-T_{jl}+\frac 
12Tg_{jl}\right] _{(1)}n^{[0}\overline{m}^{j]}n^{[0}\overline{m}^{l]} 
\nonumber
\end{eqnarray}
where $N_{(0)}=(g_{\text{kerr}}^{tt})^{-1/2}$ is the zeroth order lapse, $%
n^i,\overline{m}^j$ are two of the null vectors of the (zeroth order) tetrad
(see Ref. \cite{T73}), latin indices run from 1 to 3, and the brackets are
computed to only first order (zeroth order excluded).

To obtain $\partial _t$$\psi _4,$ the other relevant quantity in order to
start the integration of the Teukolsky equation, we can operate with $%
\widehat{\partial }_0$ on $\psi _4$ given by Eq. (\ref{psi}) to find

\begin{eqnarray}
\partial _t\psi _4 &=&N_{(0)}^\phi \partial _\phi \left( \psi _4\right) -n^i%
\overline{m}^jn^k\overline{m}^l\left[ \widehat{\partial }_0R_{ijkl}\right]
_{(1)}  \label{psipunto} \\
&&+4N_{(0)}n^{[0}\overline{m}^{j]}n^k\overline{m}^l\left[ \widehat{\partial }%
_0K_{j[k,l]}+\widehat{\partial }_0\Gamma _{j[k}^pK_{l]p}+\stackrel{\_}{%
\Gamma }_{j[k}^p\widehat{\partial }_0K_{l]p}\right] _{(1)}  \nonumber \\
&&-4N_{(0)}^2n^{[0}\overline{m}^{j]}n^{[0}\overline{m}^{l]}\left[ \widehat{%
\partial }_0\stackrel{\_}{R}_{jl}-2K_{(l}^p\widehat{\partial }%
_0K_{j)p}-2N_{(0)}K_{jp}K_q^pK_l^q\right.  \nonumber \\
&&\left. +K_{jl}\widehat{\partial }_0K+K\widehat{\partial }_0K_{jl}-\widehat{%
\partial }_0T_{jl}+\frac 12g_{jl}T-N_{(0)}TK_{jl}\right] _{(1)}  \nonumber
\end{eqnarray}
where we made use of the equality 
\[
g_{ip}\widehat{\partial }_0g^{pj}=2NK_i^j. 
\]

The derivatives appearing in Eq. (\ref{psipunto}) can be obtained from Eq. (%
\ref{r00}) 
\begin{equation}
\widehat{\partial }_0K=N_{(0)}K_{pq}K^{pq}-\stackrel{\_}{\nabla }%
^2N_{(0)}-N_{(0)}^{-1}T_{00},  \label{Kpunto}
\end{equation}
from Eq. (\ref{G00}) 
\begin{equation}
\widehat{\partial }_0\stackrel{\_}{R}=2K^{pq}\widehat{\partial }%
_0K_{pq}+4N_{(0)}K_{pq}K_s^pK^{sq}-2K\widehat{\partial }_0K-2\widehat{%
\partial }_0T_0^0,  \label{Rpunto}
\end{equation}
and from Eqs. (\ref{r3ijkl}) and (\ref{curvextr}) 
\begin{eqnarray}
\widehat{\partial }_0R_{ijkl} &=&-4N_{(0)}\left\{ K_{i[k}\stackrel{\_}{R}%
_{l]j}-K_{j[k}\stackrel{\_}{R}_{l]i}-\frac 12\stackrel{\_}{R}\left(
K_{i[k}g_{l]j}-K_{j[k}g_{l]i}\right) \right\}  \label{Rijklpunto} \\
&&\ +2g_{i[k}\widehat{\partial }_0\stackrel{\_}{R}_{l]j}-2g_{j[k}\widehat{%
\partial }_0\stackrel{\_}{R}_{l]i}-g_{i[k}g_{l]j}\widehat{\partial }_0%
\stackrel{\_}{R}+2K_{i[k}\widehat{\partial }_0K_{l]j}-2K_{j[k}\widehat{%
\partial }_0K_{l]i}.  \nonumber
\end{eqnarray}

Note that in the last three equations we have taken explicitly the lapse to
the zeroth perturbative order. This is so because in building up $\partial
_t $$\psi _4$ explicitly all dependence on $N_{(1)}$ cancels out. To prove
this one can do the explicit calculation for the Kerr background using
computer algebra. An alternative is to notice that $\partial_0 \psi_4= {\cal %
L}_t \psi_4$ where $t^a$ is a vector that includes the background and first
order perturbations of the lapse and shift. If one now expands out this
expression one gets $\partial_0 \psi_4= {\cal L}_{t_{(0)}}\psi_{4_{(0)}}+%
{\cal L}_{t_{(0)}}\psi_{4_{(1)}} +{\cal L}_{t_{(1)}}\psi_{4_{(0)}}$. Now,
since $\psi_{4_{(0)}}$ vanishes identically for all time, the only
contribution one has is $\partial_0 \psi_4 = {\cal L}_{t_{(0)}}%
\psi_{4_{(1)}} $. Therefore the time derivative of $\psi_4$ does not depend
on the perturbative lapse and shift, since neither ${\cal L}_{t_{(0)}}$ (by
construction) nor $\psi_{4_{(1)}}$ (due to the proof we gave above), do.

The other pieces needed to build up $\partial _t$$\psi _4$ only out of
hypersurface data are $\widehat{\partial }_0K_{ij},\widehat{\partial }%
_0\Gamma _{ij}^k,$ and $\widehat{ \partial }_0\stackrel{\_}{R}_{ij}$ that
are given by Eqs. (\ref{rij}), (\ref{gamapunto}) and (\ref{Rijpunto})
respectively. As before, we have to consider the zeroth order lapse only,
for instance 
\begin{equation}
\widehat{\partial }_0K_{ij}=N_{(0)}\left[ \bar R%
_{ij}+KK_{ij}-2K_{ip}K^p{}_j-N_{(0)}^{-1}\bar \nabla _i\bar \nabla
_jN_{(0)}-T_{ij}+\frac 12Tg_{ij}\right] _{(1)}.  \label{Kijpunto}
\end{equation}

This completes our proof. A check of the relations (\ref{psi}) and (\ref
{psipunto}) can be made in the Schwarzschild background for close limit
initial data where \cite{CKL98} at t=0 we have $\partial _t\psi =-\frac{2M}{%
r^2}\psi .$

\section{Discussion}

The issue of expressing $\psi $ explicitly in terms of hypersurface data
only appears as of a purely technical character, but it is of great
practical use. Especially when one thinks of the important role played by
first order perturbations as testbeds for comparison with full numerical
integration of Einstein equations. Note that since Eqs. (\ref{psi}) and (\ref
{psipunto}) hold on any $t=$ constant slice of the space time can not only
be used to build up initial values for $\psi $ and $\partial _t\psi $, but
also at a later time to extract fully numerically generated wave forms.

The above equations provides the desired link between initial data
(consisting of $\stackrel{\_}{g}_{ij}$and $K_{ij}$ ) and the Weyl scalar $%
\psi _4$. Geometrical objects like $\stackrel{\_}{\Gamma }_{ij}^k,$ $\bar R%
_{ij}$ and $\bar R_{ijkl}$ involve first and second derivatives of the
metric. Since astrophysical initial data for Kerr perturbations are
numerically generated \cite{BP98} this fact has to be taken into account.
Expression (\ref{psi}) also includes a source term that allows to
incorporate perturbations generated by particles or accretion disks around
Kerr holes.

If one chooses to work in the Teukolsky equation with $\psi _0=-C_{\alpha
\beta \gamma \delta }l^\alpha m^\beta l^\gamma m^\delta $ , which gives a
better representation of ingoing gravitational waves, a completely analogous
procedure applies to connect it to hypersurface data upon replacement of the
double contractions with the corresponding null vectors $l^\alpha $ and $%
m^\beta $ instead of $n^\alpha $ and $\stackrel{\_\_}{m}^\beta .$

Finally, we have been able to write $\psi _4$ and $\psi _0$ on the
hypersurface $\Sigma _t,$ but we did not said why. In fact it is not
warranted that one can do that with any object defined on the spacetime. Is
this because they are first order gauge invariant objects? This shouldn't be
enough since we checked that for $\psi _3$ (and the same for $\psi _1$), we
do not succeed in writing them in terms only of objects on the slice $t=$
constant. The key point here seems to be that $\psi _4$ and $\psi _0$ are
also invariant under tetrad rotations and then directly connected to
physical quantities, while $\psi _3$ and $\psi _1$ are not.

\begin{acknowledgments}
The authors thank A.Ashtekar and W.Krivan for useful discussions.
C.O.L. is a member of the Carrera del Investigador Cient\'\i fico of CONICET,
 Argentina
and thanks FUNDACI\'ON ANTORCHAS for partial financial support.
This work was supported by Grant NSF-PHY-9423950, by funds of the Pennsylvania
State University and its office for Minority Faculty Development, and the
Eberly Family Research Fund at Penn State. JP also acknowledges support
form the Alfred P. Sloan Foundation.
\end{acknowledgments}

\appendix 

\section{Alternative equations}

We can put all this together to yield the following expression of the
first order perturbation in $\psi_4$ in terms of perturbations in the
3-metric $\delta g_{ij}$, perturbations in the extrinsic curvature
$\delta K_{ij}$, and several quantites from the the background (Kerr)
geometry, the spatial metric $\ ^{(3)} {g^{(0)}}_{ij}$, the extrinsic
curvature $K^{(0)}_{ij}$, the lapse $N^{(0)}$
and the shift $N^{(0)}_i$. We have already
argued that first order perturbations of the principal null vectors
$n^\mu$ and $\stackrel{\_\_}{m}^\mu$
will not contribute to $\delta \psi_4$ so we have
$$ \delta{\psi_4 }= {\delta A_{ijkl}}{n^i}{{\bar m}^j}{n^k}{{\bar m}^l} 
	+2 {\delta B_{ijk}} {n^j}{{\bar m}^k}
		[{n^0}{{\bar m}^i}-{n^i}{{\bar m}^0}]
	 + {\delta C_{ij}}[{n^0}{{\bar m}^i}{n^0}{{\bar m}^j}
		+{n^i}{{\bar m}^0}{n^j}{{\bar m}^0}
		-{n^0}{{\bar m}^i}{n^j}{{\bar m}^0}
		-{n^0}{{\bar m}^j}{n^i}{{\bar m}^0}]
$$

where
\begin{eqnarray*}
 {\delta A_{ijkl}}&=& \delta{\ ^{(3)} R_{ijkl}}+
	[ {K^{(0)}_{jl}}\delta {K_{ik}}+{K^{(0)}_{ik}}{\delta {K_{jl}}}
	-(k\leftrightarrow l)]\\
{\delta B_{ijk}}&=&{N^{(0)}}[D_j\,\delta{K_{ik}}
	 	-{1\over2}[D_k\,\delta {\ ^{(3)} g_{mi}}+
		D_i\,\delta{\ ^{(3)} g_{mk}}-D_m\,\delta{\ ^{(3)} g_{ik}}]
	{\ ^{(3)} {g^{(0)}}^{lm}}{K^{(0)}_{lj}}- (k\leftrightarrow j)]\\
 & &	+ {N^{(0)l}} {\delta A_{lijk}}
		+A^{(0)}_{lijk}{\delta^{(3)} g^{lm}}{N^{(0)}}_m\\
\delta {C_{ij}}&=&{N^{(0)2}} {A^{(0)}_{iljm}}\delta{\ ^{(3)} g^{lm}}
	+{N^{(0)2}}{\delta A_{iljm}}{\ ^{(3)} {g^{(0)}}^{lm}}
	- [\delta{B_{ijl}}{N^{(0)l}}+{B^{(0)}_{ijl}}\delta^{(3)} g^{lm}
 {N^{(0)},}_{m}+{A^{(0)}_{jil}}\delta {\ ^{(3)} g}^{lm} {N^{(0)}}_{m}\\
& &+\delta{A_{jil}}{N^{(0)l}}+\delta{A_{iljm}}{N^{(0)l}}{N^{(0)m}}
	+{A^{(0)}}_{iljm}{N^{(0)},}_{k}\delta^{(3)} g^{kl}{N^{(0)}}^{m}
	+{A^{(0)}}_{iljm}{N^{(0)},}^{l}\delta^{(3)} g^{km}{N^{(0)}}_{k}]
\end{eqnarray*}
and
$$
\delta{\ ^{(3)} R^{i}_{jkl}}=
{1\over2}D_k[{\ ^{(3)}{g^{(0)}}^{im}}({D_l\,\delta^{(3)} g_{mj}}
	+{D_j\,\delta^{(3)} g_{ml}}-{D_m \delta^{(3)} g_{jl}})]
	 - (k\leftrightarrow l)
$$
To calculate $\partial_t {\psi_4}$
we use the above expression for $\delta\psi_4$
and plug in $\partial_t\delta^{(3)}{g}_{ij}$ and $\delta\partial_t K_{ij}$
for $\delta^{(3)} g_{ij}$ and 
$\delta K_{ij}$ in the above, respectively. Where,
$\partial_t\delta^{(3)}{ g}_{ij}$
and $\delta\partial_t K_{ij}$ can be obtained from Einstein's
equations as follows:
\begin{eqnarray*}
\partial_t\delta^{(3)}{ g}_{ij}&=& -2{N^{(0)}} \delta K_{ij} 
	+ {N^{(0)}}^{k} \delta^{(3)} g_{ij,k}+{N^{(0)}}_{l}
		 \delta^{(3)} g^{lk} {\ ^{(3)}{g^{(0)}}}_{ij,k} 
	+ \delta^{(3)} g_{ik} {N^{(0)k}}_{,j}\\
 & &	+{\ ^{(3)} {g^{(0)}}}_{il}[\delta^{(3)} g^{kl}{N^{(0)}}_{k}]_{,j}
	+{\ ^{(3)} {g^{(0)}}}_{lj}[\delta^{(3)} g^{kl}{N^{(0)}}_{k}]_{,i}
	+\delta^{(3)} g_{kj} {N^{(0)k}}_{,i}\\
\delta\partial_t K_{ij}&=&{1\over2}
[D_j\,\delta^{(3)} g_{mi}+D_i\,\delta^{(3)}{g_{mj}}
	-D_m\,\delta^{(3)}{g_{ij}}]{\ ^{(3)} {g^{(0)}}}^{mk}{N^{(0)}}_{,k}\\
 & &	+{N^{(0)}}[\delta^{(3)} R_{ij} - 2 {K^{(0)k}}_{j}\delta K_{ik}
	-2\delta{K^k}_{j}{K^{(0)}}_{ik}+{K^{(0)}}_{ij} \delta K 
	+{K^{(0)}} \delta K_{ij}]\\
 &  &	+ {N^{(0)}}^{k} \delta K_{ij,k} + \delta K_{ik} {N^{(0)k}}_{,j}
	+ \delta K_{kj} {N^{(0)k}}_{,i}
	+ {K^{(0)}}_{il}[\delta^{(3)} g^{kl}{N^{(0)}}_{k}]_{,j}\\
 & &	+{K^{(0)}}_{lj}[\delta^{(3)} g^{kl}{N^{(0)}}_{k}]_{,i}
	+{N^{(0)}}_{l} \delta^{(3)} g^{lk} {K^{(0)}}_{ij,k}
\end{eqnarray*}
where $\delta K= {\ ^{(3)} {g^{(0)}}}^{ij}\delta K_{ij}+
{K^{(0)}_{ij}}\delta{\ ^{(3)}{g}}^{ij}$
and
$\delta {K^{i}}_{j}=\delta K_{jk}{\ ^{(3)}{g^{(0)}}}^{ki}+
{K^{(0)}}_{jk}\delta{\ ^{(3)}{g}}^
{ki}$.

\end{document}